\documentclass[twocolumn,showpacs,showkeys,preprintnumbers]{revtex4}
\usepackage{graphicx}
\usepackage{subfigure}
\usepackage{lscape}
\usepackage{dcolumn}
\usepackage{bm}
\usepackage{color}

\begin{document}

\title{Newly born extragalactic millisecond pulsars as efficient emitters of PeV neutrinos}
\author{$^1$Rajat K. Dey}
\email{rkdey2007phy@rediffmail.com}
\author{$^1$Animesh Basak}
%\email{ab.astrophysics@rediffmail.com}
\author{$^1$Sabyasachi Ray}
%\email{methesabyasachi@gmail.com}
\author{$^2$Tamal Sarkar}
%\email{methesabyasachi@gmail.com}

\affiliation{$^1$Department of Physics, University of North Bengal, Siliguri, WB 734 013 India\\
	$^2$High Energy \& Cosmic Ray Research Center, University of North Bengal, Siliguri, WB 734 013 India 
	}

\begin{abstract}
The origins of the diffuse flux of cosmogenic PeV neutrinos detected in the IceCube experiment during 2010 - 2017 remain unidentified. A population of extragalactic newly born fast spinning pulsars are investigated as possible candidates for generating the PeV energy scale  neutrinos. A two-step mechanism of particle acceleration is adopted for transferring energy from star rotation to high energy electrons. Electrons might be boosted up to $\approx 0.01$ EeV energies and above, and produce ultra-high-energy (UHE) neutrinos and gamma rays when these electrons interact with low energy positrons and soft radiations in the acceleration zone. The theoretically derived extragalactic diffuse muon neutrino flux in the energy range [1-10] PeV is found consistent with the IceCube level if only a fraction ($\eta_{k} \approx 0.31\%$) of the total bolometric luminosity of pulsars are transferred to power the PeV neutrinos. Using the above value of the cosmic ray electron loading parameter $\eta_{k}$, the diffuse gamma ray flux from inverse Compton scattering off UHE electrons in the soft radiation field might be predicted from the model.
\end{abstract}

\pacs{??? 96.50.sd, 95.75.z, 96.50.S} 
\keywords{acceleration, radiation mechanism, galaxies, pulsars, neutrinos, gamma ray}
\maketitle

\section{Introduction}
%\label{sec:intro}
The discovery of diffuse flux of ultra-high-energy (UHE) cosmogenic neutrinos, reported by IceCube in 2013, has some features that intuitively favors possible extragalactic sources, although such objects remain as yet unexplored [1]. Well-coordinated global multi-messenger astronomical observations of neutrinos, cosmic rays, electromagnetic radiation across a wide spectrum, and also gravitational waves seems a promising contemporary approach to extract crucial information on basic astrophysical issues. The neutrinos can travel unaffectedly through the densest environment from any astrophysical object. This feature makes them a unique probe that can lead an observer back to their sources. The UHE gamma rays are produced simultaneously with neutrinos, cannot escape their sources owing to absorption in their production region or are absorbed in background radiations. An association of IceCube neutrinos with UHE gamma rays would provide an opportunity of identifying the source(s) of astrophysical neutrinos.      

Everyone has no doubt about a fact that the cosmic rays/electrons we detect at an energy above $0.05$ EeV are mostly of extragalactic origin. However this does not mean that we cannot observe extragalactic neutrinos from cosmic rays/electrons at an energy below $0.05$ EeV. In fact the first extragalactic association with the blazar TXS0506+056 was
obtained with a neutrino event of $290$ TeV [2]. It has also been opined that the known galactic objects do not possess the required energetics to produce cosmic rays/electrons around $\sim 0.05$--$0.1$ eV and above. In the present scenario, it is expected that the production of PeV energy neutrinos depends strongly on the power of generated electrons from the MSPs. A transition from galactic to extragalactic origin of UHE cosmic rays or electrons is expected to occur favourably at energies around $\sim 0.1$--$1$~EeV, and they almost remain as yet unidentified sources of extragalactic origins for the UHE neutrinos. The detection of PeV energy scale neutrinos in the IceCube experiment may unfold a new possibility to correlate them with those probable extragalactic sources. It is thus seemingly advocated that the extragalactic young MSPs might be able to supply enough energy to accelerate electrons to UHEs and thereby producing PeV energy scale neutrinos and gamma rays. As these MSPs are so powerful it is therefore expected that one might have seen the effect of the ones present in our galaxy, and hence giving a possibility to observe that ones.

Different searches for astrophysical neutrinos by the IceCube experiment focus on high-energy events that start in the detector volume or that originate in the Northern Hemisphere. During 7.5 years operations (2010-2017), about five PeV neutrinos with energies in the interval [1-10] PeV have been detected [3-5]. These neutrinos might have originated from extragalactic sources of cosmic rays/electrons. These PeV energy neutrino announcement by IceCube had created huge excitement among researchers, and turning their attention towards the search of their astrophysical origin. Usually, all models consider that gamma rays are produced in $\rm{pp}$ and $\rm{p}\gamma$ hadronic interactions that would also generate PeV neutrinos within the source [5-8]. The new-born fast spinning pulsars are considered in this work as probable sources of IceCube neutrinos, and some of the observed gamma rays in several Imaging Atmospheric Cherenkov Telescopes (IACTs). In the framework of an alternative particle acceleration model, electrons being pumped to UHEs if energy diminution processes remain insignificant during the acceleration of electrons in the magnetosphere of fast spinning pulsars [10]. These electrons with energies beyond $0.01~$EeV, might generate PeV neutrinos possibly via; lepton-hadron ($e^{-} + p \rightarrow n + \nu_{e}$), and lepton-lepton ($e^{-} + e^{+} \rightarrow \nu_{l} + \tilde{\nu_{l}}$; $l=e,{\mu}, \tau$) reactions. UHE gamma rays are also produced through the more common inverse Compton scattering (ICS) process, (${e^{-} + \gamma_{\rm low}} \rightarrow {e^{-} + \gamma_{\rm high}}$). 

If the reactions ${e^{-}e^{+}_{\rm cold}}$ and ${e^{-}p_{\rm cold}}$ are the source of PeV neutrino events, there should be a supply of UHE electrons with energies all the way over $0.01~$ EeV. The ${e^{-}\gamma_{\rm low}}$ interaction would also require the same energetic electrons for producing PeV gamma rays. Such electrons could indeed be driven successfully by the new-born fast spinning millisecond pulsars (MSPs) through the Landau damping of centrifugally driven Langmuir waves [10]. The pumping of rotational energy of a pulsar into the electric field in the pulsar magnetosphere efficiently supplies the energy for growing Langmuir waves in the bulk electron-positron plasma. The excited Langmuir waves then damp on a faster local electron beam in the vicinity of the light cylinder, accelerating them to larger energies [10-11]. The acceleration mechanism is called the Langmuir-Landau-Centrifugal-Drive (LLCD), has been applied to accelerate electrons/protons in plasmas surrounding the compact objects (pulsars and active galactic nuclei (AGNs)) [12].

The $e^{-}$ and $e^{+}$ population that emanating from the pulsar may be divided into three components: a relativistic $e^{-}$--$e^{+}$ plasma, a tail of the plasma and the remnant of the primary ultra-relativistic electron beam. The set of physical parameters, like the concentration and Lorentz factor, that would characterize of these components are denoted by $n_{1},\gamma_{1}$; $n_{2},\gamma_{2}$ and $n_{b},\gamma_{b}$. In a standard system like a pulsar, a transformation of its rotation energy into energy-driven oscillations in the $e^{-}$--$e^{+}$ plasma away from the star surface is made possible via a parametric two stream instability. The entire plasma around the pulsar is assumed to be composed of so many $e^{-}$--$e^{+}$ plasmas (i.e. multi-stream: each stream with a characteristic $n$ and $\gamma$, and also a characteristic phase ($\phi$) [12]). The linear interacting dynamics of two such streams excites the parametric pumping (or two stream instability) of plasma oscillations in the magnetosphere. Eventually the phase difference between the streams emerges as the driver to augment the energy content of the $e^{-}$-$e^{+}$ plasma in the form of Langmuir waves.

In this paper, we mainly focus on the emission of UHE neutrinos contributed by the process of purely leptonic origin, particularly, via ${e^{-}e^{+}_{\rm cold}}$ interaction in extragalactic new-born MSPs. The relevant interaction ${e^{-}\gamma_{\rm low}}$, in softer radiation field zones, leading to high-energy gamma ray flux, will be included as a possible counterpart of the detected IceCube PeV neutrinos. The interaction cross section of the process ${e^{-}e^{+}_{\rm cold}}$ reaches to the level of $\sim 10^{-32}$~cm$^{2}$ [13]. In the pulsar atmosphere, the remnant of the primary electrons/positrons and the electrons/positrons in the plasma streams usually near the low end of the distribution for which the waves do not satisfy the relative phase ($\phi_{-}$) restriction (i.e. $\phi_{-}>\pi/6$ and a more detail is given in Section II), altogether act as targets for the $e^{-}+ e^{+}$ interactions. Now, for a young MSP, the electron/positron density close to the light cylinder is $\sim 9\times10^{19}(r/R)^{2}\sim 10^{18}$~cm$^{-3}$, whereas the $e^{-}$--$e^{+}$ plasma concentration, that might have not satisfied the relative phase restriction, would vary in the range; $\sim 10^{11}$--$10^{14}$~cm$^{-3}$. Hence, the density of the targets of the $e^{-}$--$e^{+}$ interaction might reach up to $\sim 10^{18}$~cm$^{-3}$. 

It should be however mentioned that the ${e^{-}p_{\rm cold}}$ interaction scenario seems highly implausible because of very weak scattering of UHE electrons on cold protons. The process has an interaction cross-section at the level of $\sim 10^{-38}~$cm$^{2}$ [14] (and references therein) that requires proton densities at the level of $\sim 10^{20}~$cm$^{-3}$ to compete with electrons synchrotron loss. The details about the targets for the ${e^{-}e^{+}_{\rm cold}}$ and ${e^{-}\gamma_{\rm low}}$ interaction processes will be mentioned later to emphasize the state of the art of this work.  

Several models were proposed in recent papers to understand the possible origins of IceCube's detected PeV neutrinos in galactic/extragalactic astrophysical sites. The well-known hadronic interactions by cosmic rays with matter and radiative background might lead to the generation of PeV astrophysical neutrinos. These sources include cores and jets of AGNs [15-16], the prompt and afterglow regions of gamma-ray bursts (GRBS) [17-18], blazars [19], milli-second newly born pulsars/magnetars (assuming pp and p$\gamma$ interactions) [9,20], microquasars [21] and supernovae [22]. Also starburst galaxies [23], large-scale structures and galaxy clusters [24] have also been found in the literature. 

The present work, however, not only involves the lepton-lepton interactions for the generation of PeV neutrinos but also exploits an acceleration mechanism that pumps efficiently the spin down energy of the MSP into the particles' energy. Our main focus is to investigate the aftermath of the acceleration era of relativistic electrons on the relatively low energy positrons and soft radiative fields available in the pulsar environment with a view to producing the observed flux of PeV neutrinos obtained by IceCube.

The rotational energy of the pulsar is the only source to accelerate the electrons. Furthermore, the pulsar's energy delivering ability is directly linked with its time period of spinning. The most important pair of measurable parameters of pulsars that distinguishes them from each other are the spin period ($P$) and period derivative ($\dot{P}$). In addition to $P\dot{P}$, the secondary parameters like spin down power ($\dot{E}$) and star's  magnetic field ($B$), derived from the timing information $P\dot{P}$ are also provided to the ephemeris for all known pulsars (gamma ray and radio pulsars). Diagrams in [25-26], the distribution of $\dot{P}$ as a function of $P$ are shown for known/discovered pulsars along with $\dot{E}$, $B$ and age ($\tau$) contours. The proposed/undiscovered new born MSPs with $\dot{P}\sim 10^{-12}$~ss$^{-1}$ in this work are assumed to be evident in the  $P\dot{P}$ diagram (region corresponding to $P\sim 1$~ms, $\dot{P}\sim 10^{-12}$~ss$^{-1}$ and $B\sim 10^{12}$~G [25]. One should however keep in mind that such type of pulsars have not yet been reported from any observation; they are only theoretically [27] or analytically and numerically [32] predicted. Moreover, the possible maximal energy that can be reached in such a MSP was estimated via the LLCD [10]. Meanwhile, the Fermi-LAT collaboration has reported GeV gamma ray excess from the Galactic Center region of Milky way [28]. On the other hand, the diffuse TeV gamma ray excess has been observed by H.E.S.S. group [29]. A new population of MSPs could be considered to be potential candidates for explaining the Galactic Center Excess (GCE) according to some recent analyses [30]. Some sort of associations of TeV gamma rays with IceCube's TeV - PeV neutrinos has been suggested in [30]. It is a fact that neutrinos do not loose energy when passing through a radiation/matter unlike the gamma rays. Till date, the physical origins of the GCE remains a subject of debate, one may thus argue that the GC regions of galaxies and also galaxy clusters/star-forming galaxies might be effective locations of activity of the proposed MSPs. We also note that a few of the neutrino events (TeV energies) detected by IceCube, are relatively closer to the GC. The remaining are at other regions of the galaxy, and at extragalactic sites as well [1,3]. Along the way, we assume that this type of high-spinning pulsar is extremely rare in galaxies.

The plan of the paper is the following. A circumstantial description of the UHE electron production exploiting the LLCD mechanism is presented, in section 2. In section 3 we estimate the diffuse flux of UHE neutrino/gamma ray from the aftermath era of UHE electrons on the appropriate targets, while in section 4, we summarize our conclusions.
%**************************************************************

\section{Electron acceleration via LLCD}
We will now provide basic elements of the LLCD mechanism that produce super-energetic electrons over the energy $0.01~$ EeV, and finally leads to PeV neutrinos and TeV reaching at Earth as described in section 1.

A compact object with a strong rotation and a strong magnetic field can induce huge electric field, of the order of billion statvolts/cm near the surface of a fast-spinning MSP. Such a field initially uproots electrons and also protons from the neutron star's surface.
The accelerated electrons suffer curvature radiation loss and the emitted photons subsequently undergo $e^{-}$--$e^{+}$ pair creation until $E_{\gamma} \geq 2m_{e}c^{2}$, and the pairs are further accelerated and emit curvature photons. This process will continue till to a distance in star's magnetosphere from the neutron star surface where the resultant pair plasma density is sufficiently high to screen out the induced electrostatic field [31]. Next, a systematic operation of an astrophysical setting would start working ({\it{e.g.}} pulsar or AGN) in the rotating magnetosphere for transferring rotational energy from the pulsar to kinetic energy of electrons. The LLCD does act exactly what one requires here to generate UHE electrons. In highly dense plasmas of concentration in the range $\sim 10^{11}$--$10^{14}$~cm$^{-3}$ surrounding the fast spinning MSPs, the LLCD undergoes through a two-step process.

In the first step, the Langmuir waves are generated by the bulk electron-positron with the relatively lower Lorentz factor region in the pulsar's magnetosphere. These excited Langmuir waves damp on 
a local electron beam in the high Lorentz factor end of the plasma distribution, accelerating them to much higher level. This constitutes the last step, and is known as Landau damping which is a consequence of rapid Langmuir collapse. This above combination could supply relativistic electrons with energies up to EeV range in MSPs and AGNs [11,32-33]. 

From [34], we obtain the characteristic Landau damping rate of energetic plasmas on a relativistic local electron beam
\begin{equation}
\Gamma_{LD}=\frac{{n_{GJ}}\gamma_{b}\omega_{b}}{n_{p}{\gamma_{p}}^{2.5}},
\end{equation}

where $\rm{n}_{GJ}$, $\gamma_{b}$ and  $\omega_{b}={(\frac{4{\pi}e^{2}n_{GJ}}{m})}^{\frac{1}{2}}$ are Goldreich-Julian density, the Lorentz factor and the plasma frequency of the specific species on which the damping of electrostatic waves settles down. $\rm{n}_{p}$ and $\gamma_{p}$ are respectively the plasma number density and its Lorentz factor.

For the effectiveness of the LLCD in fast spinning pulsar's magnetosphere, typically the damping, and the instability growth rates in the bulk plasma, are of the order of $10^4$~s$^{-1}$ [12]. On the other hand, the kinematic rate, also called the pulsar's angular rate of rotation ($\Omega$) is close to $\approx 6\times 10^3$~s$^{-1}$. Hence the instability growth and damping rates are faster than the kinematic rate. In the first half of the LLCD, the centrifugal acceleration drives the electrostatic Langmuir waves consuming the central star's rotational energy via a parametric two stream instability with a growth rate [12],    
\begin{equation}
\Gamma_{GR}=\frac{\sqrt{3}}{2}{(\frac{\omega_{1}{\omega_{2}}^2}{2})}^{\frac{1}{3}}J_{\mu}(b)^{\frac{2}{3}},
\end{equation}

where $\rm{J}_{\mu}$ represents the Bessel's function and $\rm{b}={(\frac{2\rm{ck}}{\Omega})\rm{sin\phi_{-}}}$ with $\phi_{-}={(\phi_{\rm p}-\phi_{\rm e})/2}$ (here, $\phi_{\rm p}$ and $\phi_{\rm e}$ denote the initial phases of $\rm e^{+}$ and $\rm e^{-}$, $\Omega$ being the angular velocity of the pulsar, $\rm k$ denotes the wave vector). The several steps analytical treatment of the stream motion [12] in the magnetosphere finally ensures the possibility of the growth of perturbations. In the framework of the perturbation theory the perturbed stream densities involve imaginary exponent. The imaginary exponent can be expressed as a linear combination of the Bessel's functions of different orders through the Bessel identity [12]. The instability growth rate after exploiting some algebra takes an imaginary part through the Bessel's function. In Eq. (2), $\omega_{1,2}= \sqrt{8\pi{\rm{e}^2}\rm{n}_{1,2}/\rm{m}{{\gamma_{1,2}}^3}}$, ${\rm{n}_{1,2}}$ and ${\gamma_{1,2}}$ are the relativistic plasma density, the number density and the Lorentz factor respectively of the two species i.e. $\rm e^{+}$ and $\rm e^{-}$. 

We have learned already that the particles of electron-positron plasma have received pulsar's rotational energy. In the cosmic rest frame these particles start to slide along the magnetic field lines experiencing the centrifugal force in a frozen-in condition (radial velocity $\sim 0$) [35]. The reaction force $f_{r}$ relative to the laboratory frame pushes the particles towards the boundary of the pulsar's light cylinder. In the vicinity of the light cylinder surface the particles would gain the maximum energy from the pulsar's magnetosphere and is equal to the total work done by $f_{r}$ 
\begin{equation}
W_{e} \approx n_{1}f_{r}\delta{r}\delta{V},
\end{equation}        

where $\delta{r}\approx c/{\Gamma}$ is the scale distance and $\delta{V}$ is the corresponding volume within which pumping occurs [35-36]. The local beam electrons will receive energy by the above work done in the same volume. If $\epsilon_{e}$ is the energy gained by each beam electron, then $W_{b}\approx n_{GJ}\epsilon_{e}\delta{V}$, would therefore give the total energy gained by all the beam electrons available in the volume. The total energy gained by a beam electron has been found by making, $W_{b} \approx W_{e}$, as [11,35]   
\begin{equation}
\epsilon_{e} \approx \frac{n_{1}f_{r}\delta{r}}{n_{GJ}}.
\end{equation}

In Eq. (4), we can use, $f_{r}\approx 2{\rm mc}\Omega{(1-\Omega^{2}r^{2}/{\rm c^2})^{-3/2}}\equiv 2{\rm mc}\Omega{\xi^{-3}}$; $\xi$ is called the time lapse function and is estimated to $\sim 10^{-3}$ [34]. For the young MSP, we take $\rm P\sim 10^{-3}$ s and ${\rm{B}_{\rm{lc}}}\approx \rm{B}{(\rm{R/R_{lc}})^{3}}$ with $\rm{R}_{lc}\equiv {\rm c}/\Omega$. Here, $B_{lc}$ is the magnetic field on the light cylinder zone, $R$ and $R_{lc}$ are radii of the pulsar and its light cylinder. For the two$-$stream instability, the instability growth and Landau damping rates are large and comparable. By considering $\rm{B}\approx 10^{12}$ G, the magnetic field near the star's surface, and $\dot{\rm P}\sim 10^{-12}$ ss$^{-1}$, and for the combination; $\gamma_1\approx 1.8\times 10^5$, $\gamma_2\approx 8\times 10^5$, and $\gamma_{\rm b}\approx 7.5\times 10^7$, the desired condition i.e. $\Gamma_{GR}\sim \Gamma_{LD}\equiv \Gamma \sim 10^{4}~$s$^{-1}$ can be achieved by the LLCD [11]. In regions close to the light cylinder, the Goldreich-Julian number density $n_{GJ} \equiv (B/Pec)$ is $\approx 1.8\times 10^{12}$~cm$^{-3}$. The radius $\rm r$ of an electron from the center of the pulsar (Pulsar's own radius is taken to $\rm R\approx 10^6$ cm) is taken roughly equal to $\rm {r} \approx 5{\rm{R}}$, and for typical magnetospheric parameters of the pulsars under the present study, the Eq. (4) estimates electrons energy close to 0.27 EeV, provided, $\rm{f}_{r} \approx 3.4\times{10^{-4}}$ in CGS unit. One of the magnetospheric parameters like the stream density $\rm{n}_1$, is taken as $\approx 7.5\times 10^{14}$ cm$^{-3}$. This value for the stream density has been computed using a rough equipartition of energy in the two constituents as, $\rm{n}_1\gamma_1\approx {\rm{n}_{GJ}}{\gamma_{\rm b}}$.

The UHE electrons interact with low-energy/cold positrons beyond the light cylinder zone, the unstable $\rm Z$ boson state may form which then decays into five individual channels with different branching fractions. At $\rm Z$ boson peak, the branching ratio (BR) of the $\rm Z$ boson decay into $\nu_{l}\tilde{\nu}_{l} (l=e,\mu,\tau)$ together is roughly $\frac{1}{5}$. The remaining BR of amount $\frac{4}{5}$ accounts for the production of charged lepton pairs ($l=e^{\pm},\mu^{\pm},\tau^{\pm}$). The average percentage of UHE electrons energy carried out by the neutrinos via the unstable $\rm Z$ boson state is $\sim 20\%$. Hence, each of the neutrino, irrespective of their flavors could receive $\sim 3\%$ of the projectile's energy via the $e^{-}e^{+} \rightarrow \nu_{l}\tilde{\nu_{l}}$ reaction channel.  
\begin{equation}
E_{\nu} \approx 0.03{E_{e}}\approx 1.5(PeV)~\epsilon_{e,17}[2/(1+z)].
\end{equation} 

Here, $\epsilon_{\rm e}=\epsilon_{e,17}{(10^{17} eV)}$, being the electron energy in the cosmic rest frame and $z$ is the gravitational redshift of the source.

From [37], it was concluded that at the beginning of electron acceleration, the cooling time scale due to ICS is longer than the acceleration time scale which means insignificant loss of electron energy. In the vicinity of the light cylinder ($\rm{r\leq R_{lc}}$), the rotational energy gain of electrons is strongly limited by the mechanism called {\lq{breakdown of the bead on the wire}\rq} (BBW) approximation [38]. Beyond the light cylinder i.e. ($\rm{r > R_{lc}}$), efficient energy transferring to electrons materializes via Langmuir collapse [12]. The resulting Lorentz factor of the electrons reaches to a maximum value $\approx 10^6$--$10^7$. The soft photons in nearby regions will attain a maximum energy via ICS by the UHE electrons with average Lorentz factor $\approx 5\times 10^6$ [37].   
\begin{equation}
E_{\gamma}\approx \frac{\gamma_{max}^{2}}{10^{10}}\approx{2.5}PeV
\approx 0.025{E_{e}}\approx 1.25(PeV)\epsilon_{e,17}[\frac{2}{(1+z)}]. 
\end{equation}   

A very young MSP with rotational period $\rm{P}$, could have an associated rotational energy $\rm{E_{rot}=\frac{1}{2}I{\Omega}^{2}}$. With the the star's moment of inertia, $\rm{I}\approx 10^{45}$ g cm$^2$ and $\rm{P}\sim 1$ ms, the rotational energy takes the value $\rm{E_{rot}\approx 1.6\times 10^{52}}$ erg. Consideration of $\rm{P}\sim 1$ ms as the early phase rotational period, is appropriate to $\rm{E_{rot}\approx 10^{52}}$ erg in the present case. We account the star's spin-down in contributing to a spin-down luminosity ($\rm{L_{sd}}$). The $\rm{L_{sd}}$ is roughly equal to the electromagnetic luminosity of the MSP (assuming gravitational wave losses by the MSP is unsubstantial). Therefore the bolometric luminosity ($\rm{L_{b}}$) of the MSP then reads [39]
\begin{equation}
L_{b} \approx  L_{sd}\approx \frac{\mu^{2}\Omega^{4}}{c^3}\approx 5\times 10^{43}~erg~s^{-1},
\end{equation} 
	
where $\mu=\rm{BR^{3}}$ is the pulsar's magnetic moment under consideration. The parameters $\rm B$ and $\rm R$ are respectively the pulsar's magnetic field strength at its surface and radius. Suppose, a fraction $\eta_{k}$ is consumed in the acceleration process of electrons close to EeV energies. Electrons with an energy given below, finally impinge upon their respective targets just outside the pulsar's light cylinder.
\begin{equation}
\epsilon_{e} \approx  \frac{\eta_{k}L_{b}}{4\pi\eta_{p}R_{lc}^{2}n_{GJ}c},
\end{equation} 
	
where $\eta_{p}$ accounts the fraction of electrons participated in the acceleration process which is expected to be smaller than 1. We have already found that the Eq. (4) yielded electrons energy $\approx 0.27~$EeV, for a set of typical values of the parameters involved in the equation. Comparing Eq. (4) with Eq. (8), an estimate of the ratio $\frac{\eta_{k}}{\eta_{p}}$ in (8), can be known, and it is $\sim 60$. In the following section, it would be possible to limit the parameter $\eta_{k}$ from the comparison between the IceCube estimated and our calculated PeV neutrinos fluxes with equal contribution from all neutrino flavors.
	
It is noteworthy to mention that the electrons kinetic (flux) luminosity $\rm{L_{k,e}}$ contains a fraction of the $\rm{L_{sd}}$ or $\rm{L_{b}}$. It is therefore inevitable that, $\rm{L_{k,e}}< \rm{L_{b}}$, and the ratio of $\frac{L_{k,e}}{L_{b}}={\eta_{k}}~$, known as  the loading factor of cosmic ray electrons, has been introduced already in Eq. (8), should be definitely smaller than 1. Again the probability of $\rm e^{-}e^{+}$ and $\rm e^{-}\gamma$ interactions would alter the efficiency of conversion of the $\rm{L_{k,e}}$ to generate PeV neutrinos and gamma rays. Thus, one more parameter ($\xi_{s}$) is needed to be introduced in order to account the amount of suppression of the UHE electron flux to power UHE neutrinos and gamma rays. If the bolometric luminosity rises, the kinetic luminosity will accordingly modify the energy of emitted electrons, and hence to the neutrino/gamma ray energies [40]. 
	
In the pulsar's magnetosphere, the regions far away from the central star where the initial electrostatic field is screened, the electron-positron plasma and soft photons are sufficiently dense with the order of magnitudes $\sim 10^{11}$--$10^{14}~$cm$^{-3}$ and $\sim 10^{19}~$cm$^{-3}~$[9] respectively. Copious thermal photons emitted from the star are also available there. In the LLCD model, the $e^{-}$--$e^{+}$ plasma was assumed to be multi-stream. The plasma can be described by a widespread distribution function containing several streams, each characterized by a Lorentz factor. Langmuir waves generated by two streams when satisfy a phase difference $\phi_{-}\leq \pi/6$, can only contribute to particle acceleration [11]. Moreover, these waves possess a phase velocity which is asymptotically close to the speed of light. Hence, particles ($e^{+}_{cold}$) in the plasma streams usually near the low end of the distribution and for which the waves do not satisfy the above relative phase restriction (with $\phi_{-}> \pi/6$), may act as targets for the $e^{-}$--$e^{+}_{\rm cold}$ interactions. A certain fraction of these particles in multi-stream  may also undergo synchrotron losses and supply photons. These nascent photons and the star's ambient radiation field together may act as $\gamma_{\rm low}$, and would participate in the ICS process with the UHE electrons. 
	
It has been already stated that the parametric pumping of Langmuir waves is a highly efficient process. In the present astrophysical settings the second step i.e. Langmuir collapse is also a rapidly energy transferring process. We will now look upon very briefly about the possible energy loss mechanisms that may impose significant constraints, if any, during the energy transfer stage to electrons [11]. 
	
The overall acceleration time-scale is much smaller than the cooling time-scale for a broad range of $\gamma$s; the instability is indeed very efficient (acceleration/instability time-scale is $\sim 0.1$~ms, smaller than the Compton cooling time, $t_{cc}\sim \epsilon_{e}/P_{KN}$ due to the ICS process in the Klein-Nishina regime) [10-11]. Moreover, the cooling time-scale of the ICS process is a continuously increasing (or slow process) function of $\epsilon_{\rm e}$. The most potential synchrotron loss mechanism does not affect the continuous energy acquiring mode of electrons. The relativistic electrons leaving the pulsar's vacuum gap experience an efficient synchrotron cooling at a very short time scale $\sim 10^{-21}$~s [10]. As soon as the electrons crossing the gap, they radiate their transverse momentum, but in no time they drifting along the field lines, and reaching near the light cylinder zone. This is the space, where the Langmuir waves, always propagating along the field lines, are excited due to the event of wave interaction with particles, thereby readily suppressing the synchrotron cooling. The next possible energy loss process is the curvature radiation, the cooling time-scale of the mechanism in this environment takes much higher values than the overall acceleration time-scale and, hence does not interfere notably with the energy transfer process. The energy density of electrons with energies close to PeV and above exceeds the magnetic energy density by several orders of magnitude. At this situation, electrons follow practically straight line trajectories, thereby excluding the curvature loss from the interference with the wave energy transfer via LLCD.

\section{Diffuse neutrino and gamma ray fluxes}

The UHE electrons emanating from the pulsar interact with the cold electrons and radiative fields in regions just beyond the light cylinder, leading to the generation of UHE neutrinos and gamma rays. Each $\rm{e^{-}e^{+}}~$ interaction leads to the production of UHE neutrino-antineutrino pairs of three different flavors. According to the Eq. (5), each neutrino will then carry an amount of energy, $\approx\epsilon_{e,17}\frac{3}{(1+z)}$ PeV while leaving the young MSP with a redshift $\rm z$. On the other hand, a gamma ray photon might receive $\approx\epsilon_{e,17}\frac{2.5}{(1+z)}$ PeV energy from an ICS process. Let us introduce a parameter ($\chi_{f}$), which accounts the fraction of the UHE electrons that would successfully complete independently the $\rm{e^{-}e^{+}}~$ and ICS interactions.  

The UHE neutrino/gamma ray flux produced in $\rm{e^{-}e^{+}}~$ and ICS  processes, can be estimated from the theory. The observed neutrino flux ($\rm{{{E_{\nu}}^{2}}\Phi_{{\nu},{\tilde{\nu}}}}$) obtained by the IceCube experiment was close to the level of $10^{-9}$ to $10^{-8}$ GeV cm$^{-2}$s$^{-1}$sr$^{-1}$ in the energy bin $\sim 1$--$10$ PeV [4,41]. The UHE electron flux coming out of an extragalactic new-born MSP in terms of cosmic scale factors $\rm R$ in the Robertson-Walker (RW) metric (in the Friedmann or FRW cosmology) [42] is given by
\begin{equation}
\Phi_{e} = \frac{L_{k,e}R^{2}(t_{1})}{4{\pi}R^{4}(t_{0}){r_{1}}^2}\equiv \frac{\chi_{f}{\xi_{s}}{\eta_{k}}L_{b}R^{2}(t_{1})}{4{\pi}R^{4}(t_{0}){r_{1}}^2},
\end{equation}

where $\rm t_0$  measures the present moment i.e. the time when the UHE electrons reached at observer location. Also, the parameter $\rm t_1$ is the time when these electrons left the star, and $\rm r_1$ is the corresponding radial distance of the source at that moment. Here, $R(t_{0})$ and $R(t_{1})$ are called the present and past cosmic scale factors of the universe. 

The redshift parameter is conventionally expressed in terms of the ratio between the scale factors as [42],
\begin{equation}
z = \frac{R(t_0)}{R(t_1)}-1.
\end{equation}

Inserting Eq.(10) in (9) and finally rearranging parameters in it, Eq.(9) leads to the following
\begin{equation}
\Phi_{e}(L_{b}) = \frac{{\chi_{f}}{\xi_{s}}{\eta_{k}}L_{b}}{4{\pi}R^{2}(t_{0}){r_{1}}^2{(1+z)}^2}
\end{equation}

We will now introduce a function $\rm f(t_{1}/z,L_{b})$ that actually gives the number density of MSPs in a luminosity range in the universe. Then, $\rm f(z,L_{b})dL_{b}$ accounts the number of sources per unit volume with luminosities between $\rm L_{b}$ and $\rm L_{b}+dL_{b}$ at a redshift $z$ [42]. It was theoretically suggested that these type of MSPs might have formed at distant galaxies [27, 45-44]. The function $\rm f(t_{1}/z,L_{b})$ is derived from a best-fit luminosity function to  galaxy survey X-ray data including the cosmological evolution. The amount of UHE flux of electrons contributed by the MSPs from distances in the range, $\rm r_{1}:r_{1}+dr_{1}$ with luminosities between $\rm L_{b}:L_{b}+dL_{b}$ leads to    
\begin{equation}
d\Phi_{e,d} = 4\pi\Phi_{e}(L_{b})R^{2}(t_{1}){r_{1}}^{2}f(t_{1}/z,L_{b})|dt_{1}|\frac{dL_{b}}{L_{*}},
\end{equation}

where $L_{*}$ is called the break luminosity.

The analysis of {\it{Chandra}} Deep Field North (CDFN) [45-46] and South (CDFS) [47] X-ray surveys ($>2$~keV) found that the X-ray luminosity function (XLF) of active galaxies are parametrized by the luminosity-dependent density evolution (LDDE) model [48-50]. More specifically, an XLF is the number of active galaxies per unit volume with X-ray luminosities in a range. The luminosity-dependent density evolution (LDDE) model introduces a well represented XLF at a given $z$ by combining a double power-law luminosity function (LF) with a luminosity-dependent evolution term in the following form [51-52]
\begin{equation}
f(z,L_{b}) = \frac{A_{*}}{[(\frac{L_{b}}{L_{*}})^{\gamma_{1}}+(\frac{L_{b}}{L_{*}})^{\gamma_{2}}]}{e(z,L_{b})},
\end{equation} 

where $\gamma_1$ and $\gamma_2$ are slopes below and above the break luminosity $L_{*}$. In the LDDE model, the evolution term of the XLF is given by 
\begin{equation}
e(z,L_{b})=\left\{\begin{array}{lll}
(1+z)^{p_1};\       [z\leq z_{c1}(L_{b})] \\
(1+z_{c1})^{p_1}{(\frac{1+z}{1+z_{c1}})}^{p_2};\      [z_{c1}(L_{b})<z\leq z_{c2}]\\
(1+z_{c1})^{p_1} {(\frac{1+z_{c2}}{1+z_{c1}})}^{p_2}{(\frac{1+z}{1+z_{c2}})}^{p_3};\     [z>z_{c2}]
\end{array}
\right. 
\end{equation}

where $p_1$ to $p_2$ represent the evolution index range for the first cut-off redshift $z_{c1}$, while the cut-off $z_{c2}$ follows the index change from $p_2$ to $p_3$. A redshift cut-off is a value of $z$ at which the luminosity evolution term changes signs. The luminosity dependence of the index $p_1$ is expressed as [53],
\begin{equation}
p_{1}(L_{b}) = p_{1}^{*}+\beta_{1}(log{L_{b}}-log{L_{p}}).
\end{equation} 

The luminosity dependent cut-off redshifts $z_{c1}(L_{b})$ and $z_{c2}(L_{b})$ below and above the luminosity thresholds of $L_{a1}$ and $L_{a2}$ are  
\begin{equation}
z_{c1}(L_{b})=\left\{\begin{array}{ll}
z_{c1}^{*}(\frac{L_{b}}{L_{a1}})^{\alpha_{1}};\      [L_{b}\leq L_{a1}] \\
z_{c1}^{*};\      [L_{b} > L_{a1}]
\end{array}
\right. 
\end{equation}
\begin{equation}
z_{c2}(L_{b})=\left\{\begin{array}{ll}
z_{c2}^{*}(\frac{L_{b}}{L_{a2}})^{\alpha_{2}};\      [L_{b}\leq L_{a2}] \\
z_{c2}^{*};\      [L_{b} > L_{a2}]
\end{array}
\right.
\end{equation}

The several best-fit model parameters over the  redshift range $0.002$--$5$ from the maximum likelihood estimation of the hard X-ray CDF survey data for the XlF employing the LDDE model are listed in Table 1 [51,54]. The Eqs. (5) and (8) suggest that more luminous pulsars produce more energetic electrons and hence neutrinos. IceCube's detected PeV neutrinos correspond to electron energies in a certain interval and those energetic electrons require bolometric luminosity in the range roughly $logL:42$--$44.5$. To obtain the luminosity dependent p1 parameter from X-ray data, a linear fitting was carried out in the range $logL:42$--$44.5$.

\begin{table*}
	\begin{center}
		\begin{tabular}
			{|l|r|} \hline
			
			{\rm{Parameters}}& {\rm{Values}}\\ \hline
			
			$A_{*}$& $3.20^{c}$\\ \hline 
			$\gamma_{1}$& $0.96$\\ \hline 
			$\gamma_{2}$& $2.71$\\ \hline
			$p_{1}$& $3.71$\\ \hline
			$p_{2}$& $-1.5$\\ \hline
			$p_{3}$& $-6.2$\\ \hline 
			$p_{1}^{*}$& $4.78$\\ \hline
			$z_{c1}^{*}$& $1.86$\\ \hline
			$z_{c2}^{*}$& $3.0$\\ \hline
			$\alpha_{1}$& $0.29$\\ \hline
			$\alpha_{2}$& $-0.1$\\ \hline
			$\beta_{1}$& $0.84$\\ \hline
			$logL_{*}$& $43.97$\\ \hline
			$logL_{a1}$& $44.61$\\ \hline
			$logL_{a2}$& $44.0$\\ \hline
			$logL_{p}$& $44.0$\\ \hline    							
		\end{tabular} 
		\footnotesize{\item[$^{c}$] In $10^{-6}$ ~h$^3_{67.8}$~Mpc$^{-3}$ units; h$_{67.8}\approx 1$.}
		\caption {Summary of the best fitted model parameters for XLF in the (2-10) keV X-ray band [51]. The quantities $L_{*}$, $L_{a1}$, $L_{a2}$, $L_{p}$ are expressed in erg~s$^{-1}$.}
	\end{center}
\end{table*} 

We now recall Eq. (12) and use Eq. (10) to (11), and $\rm |dt_{1}|=\frac{c}{H_{0}(1+z)^{5/2}}~dz$ in (12). 
\begin{equation}
d\Phi_{e,d} = \frac{c{\chi_{f}}{\xi_{s}}{\eta_{k}}{f(z,L_{b})}{dL_{b}}dz}{(1+z)^{13/2}{H_{0}}{L_{*}}}.
\end{equation}

We can now include the XLF i.e. Eq. (13), the diffuse flux of UHE electrons can be calculated as
\begin{equation}
\Phi_{e,d} = \frac{cA_{*}{\chi_{f}}{\xi_{s}}{\eta_{k}}}{{4{\pi}H_{0}}{L_{*}}}\int{\frac{L_{b}dL_{b}}{[(\frac{L_{b}}{L_{*}})^{\gamma_{1}}+(\frac{L_{b}}{L_{*}})^{\gamma_{2}}]}\int{\frac{e(z,L_{b})dz}{(1+z)^{13/2}}}}, 
\end{equation}
in erg~cm$^{-2}$s$^{-1}$sr$^{-1}$
	
The luminosity integral is calculated numerically with the appropriate luminosity limits. From the Eq. (5), we can further ascertain that electrons generated via LLCD should gain energies at least in the range $(\epsilon_{e,17}\approx 0.34$--$20)~$ for obtaining IceCube neutrinos in the energy range $(1.004$--$10)~$ PeV. These electrons require a luminosity in the range $(L_{min}\approx 6.12\times 10^{42}:L_{max}\approx 3.6\times 10^{44})$ erg s$^{-1}$: \[\int^{L_{max}}_{L{min}}\frac{{L_{b}}{dL_{b}}}{[(\frac{L_{b}}{L_{*}})^{\gamma_{1}}+(\frac{L_{b}}{L_{*}})^{\gamma_{2}}]}\approx 1.127\times 10^{88}.\]~~in~(erg~s$^{-1}$)$^{2}$
	
The redshift integral consists of three parts following three redshift ranges: $0.002$--$ z_{c1}(L_{b})$; $z_{c1}(L_{b})$--$z_{c2}(L_{b})$; $z_{c2}(L_{b})$--$5$, with $z_{c1}(L_{b}) \approx 0.534$; $z_{c2}(L_{b})\approx 3.61$. We have used relevant parameters from the Table 1, and also Eq. (16) to (17) with $L_{b}\approx 10^{42}$ erg~s$^{-1}$ to calculate $z_{c1}(L_{b})$ and $L_{b}\approx 10^{43}$ erg~s$^{-1}$ for $z_{c2}(L_{b})$ (also followed $f$ (in ~Mpc$^{-3}$) versus $z$ curves, in [51]). Hence,
	
\begin{center}
	$I=\int{\frac{e(z,L_{b})dz}{(1+z)^{13/2}}}=I_{1}+I_{2}+I_{3},$
\end{center}	

where,\\\\
	$I_{1}=\int^{0.534}_{0.002}{\frac{(1+z)^{p_{1}}dz}{(1+z)^{13/2}}}\approx 0.294;$\\ $I_{2}=\int^{3.61}_{0.534}\frac{(1+z_{c1})^{p_{1}}{(\frac{1+z}{1+z_{c1}})}^{p_{2}}dz}{(1+z)^{13/2}}\approx 0.0663;$\\ $I_{3}=\int^{5}_{3.61}\frac{(1+z_{c1})^{p_1} {(\frac{1+z_{c2}}{1+z_{c1}})}^{p_2}{(\frac{1+z}{1+z_{c2}})}^{p_3}dz}{(1+z)^{13/2}}\approx 1.21\times 10^{-5}.$\\ 
	
The UHE flux of electrons with the above values of integrals, $I_1$, $I_2$ and $I_3$ is
\begin{equation}
\Phi_{e,d} \approx 4.06\times 10^{87}\times\frac{cA_{*}{\chi_{f}}{\xi_{s}}{\eta_{k}}}{{4\pi}{H_{0}}{L_{*}}}. 
\end{equation}
in erg~cm$^{-2}$s$^{-1}$sr$^{-1}$.

Finally, the diffuse neutrino (per flavor)/gamma ray flux contributed by the MSPss at $z=0.002$--$ 5$ with the well represented LF is 
\begin{equation}
\Phi_{(\nu_{l},\tilde{\nu}_{l})/{\gamma}} \approx 2.51\times 10^{90}\times\frac{cA_{*}{\chi_{f}}{\xi_{s}}{\eta_{k}}\chi_{o}}{{{4\pi}H_{0}}{L_{*}}}. 
\end{equation}
in GeV~cm$^{-2}$s$^{-1}$sr$^{-1}$.\\

In each interaction ($e^{-}+e^{+}$) that takes place beyond the pulsar's light cylinder zone, one muon neutrino and one muon anti-neutrino are produced (also possibility of $\nu_{e}\tilde{\nu_{e}}$ and $\nu_{\tau}\tilde{\nu_{\tau}}$ pairs production). During their propagation from the source (flavor ratio at source point, $1:1:1$) to the detector location, they suffer oscillations but flavor ratio remains unchanged. $\chi_0$ accounts the ratio between the total number of $\nu_{l}\tilde{\nu}_{l}$ at the detector location (earth) and at the source [55-56]. Hence, $\chi_{o} = 1$ is appropriate to account the effect. Here, we take Hubble's constant, $\rm H_{0}\approx 67.8~$km/s/Mpc [57-58]. We assume that $50\%$ of the electron kinetic flux could contribute to the overall neutrino and gamma ray fluxes. The remaining percentage of electrons may contribute to the UHE electron flux and the electromagnetic radiation of various wavelengths. Hence it is appropriate to set $\xi_{s}\approx 0.5$ for the estimation of neutrino/gamma ray flux. We further assume that the $\rm e^{-}e^{+}$ channel goes on $50\%$ of the reaction time and the other channel i.e.  the ICS consumes the other half. Hence we set $\chi_{f}=\frac{1}{2}$ for each mode of interaction in order to modify the UHE neutrino/gamma ray flux.   

Using Table 1 and other appropriate parameters elsewhere in the paper, the PeV neutrino flux per flavor results from the theory with the LDDE model for the XLF, is $\approx 8.44\times 10^{-7}\eta_{k}$ in GeV~cm$^{-2}$s$^{-1}$sr$^{-1}$. To obtain the diffuse gamma ray flux one needs to set $\chi_{o}$ equal to 1 as well. Hence, the theoretically derived diffuse PeV gamma ray flux at source via the ICS process would also be the same.  

The best-fit power law for the extraterrestrial IceCube neutrino flux is $\rm J_{{\nu},{\tilde{\nu}}}(E_{\nu})={E_{\nu}}^{2}\Phi_{{\nu},{\tilde{\nu}}}(E_{\nu})\approx 2.19\times 10^{-8}(\frac{E_{\nu}}{0.1~PeV})^{-0.91}~$GeV~cm$^{-2}$~s$^{-1}$~sr$^{-1}$ [59]. The energy spectra of neutrinos can be described by $\rm J_{{\nu},{\tilde{\nu}}}(E_{\nu})dE_{\nu}={E_{\nu}}dN_{{\nu},{\tilde{\nu}}}$, where $\rm dN_{{\nu},{\tilde{\nu}}}=\Phi_{{\nu},{\tilde{\nu}}}(E_{\nu})dE_{\nu}$ being the number of neutrinos in the energy interval, $\rm E_{\nu}$:$\rm E_{\nu}+dE_{\nu}$. Finally, the total observed diffuse energy flux of neutrinos per flavor at Earth in the energy range, $\approx~1$--$10$ PeV can be written following [59]

\begin{equation}
\Phi_{({\nu},{\tilde{\nu})},d}=\int^{10}_{1}{E_{\nu}}^{-1}J_{({\nu},{\tilde{\nu}})}(E_{\nu})dE_{\nu}\approx 2.59\times 10^{-9}\\
\end{equation}
in GeV cm$^{-2}$s$^{-1}$sr$^{-1}$.

The expected muon neutrino energy flux leads to give a quantitative measure on the amount of conversion of the bolometric luminosity to power the Icecube's measured PeV neutrino flux. It is found that the flux arising out of the model calculation can fulfill the IceCube limit well if $\eta_{k}\approx 0.31\%$.

In the present scenario, three different modes of interactions were proposed initially. The $e^{-}p_{cold}$ interaction was completely ruled out in the fray of neutrino production. The remaining two modes of interactions i.e. the $e^{-}e^{+}_{cold}$ and the ICS have been considered in the work. The production of PeV neutrinos and gamma rays depends strongly on the generation of UHE electrons in MSPs. Here, we have concluded that the theoretical prediction for the PeV scale neutrino flux will be consistent with IceCube's diffuse flux of neutrinos in the interval [1-10] PeV if $\eta_{k}\sim 0.31\%$ of the bolometric luminosity of MSPs was consumed to power UHE neutrinos and gamma rays together. Involvement of other parameters like $\xi_{s}$, $\chi_{f}$ and $\chi_{o}$ may fix the percentages of the bolometric luminosities required independently to power UHE neutrinos and gamma rays. It was however being noticed that the percentage of the required bolometric luminosity would vary depending upon the consideration of different XLFs with either cosmological evolution or not. For AGNs, different choices of the LFs yielded different percentages of the measured luminosity [8,60].

The produced PeV gamma rays leaving their emission regions, are unlikely to reach at earth with the same energy and flux. The energy and flux of a major fraction of PeV gamma rays are expected to be modified because of strong absorption of these gamma rays by the well spread extragalactic background radiations and also their cosmological distances . UHE gamma rays undergo $\gamma\gamma$ absorption due to $e^{+}e^{-}$ generation on cosmic microwave background (CMB) photons. They may also interact with galactic and extragalactic radio background photons. For gamma rays with energies, $\geq 1~$PeV, the interaction mean free path due to $e^{+}e^{-}$ production on CMB is somewhat $\geq 10$ kpc [61]. In agreement with the above conjecture, no extragalactic PeV gamma rays are going to be detected in future [62]. However, a very marginal fraction of the PeV gamma rays with energies, $1 \leq \rm E_{\gamma} < 2.5$~PeV) from low $z \approx 0.002$--$1$ extragalactic pulsars (nearby pulsars) might reach the Earth. It should be however mentioned that the Tibet AS$_{\gamma}$ Collaboration has reported the detection of PeV gamma rays at Earth which are coming out from the galactic disk in a recent work [63]. In their work, the produced PeV gamma rays might suffer very negligible absorption over the distance between the galactic disk region and the Earth ($\sim$ few kpc), and hence could reach the detector.

\section{Summary and conclusion}

We have theoretically derived the diffuse flux of PeV neutrinos ever detected through a specific lepton-lepton interaction channel in newly born extragalactic MSPs. A possibility of the generation of PeV gamma rays through the ICS process has also been investigated from the theory. Our conclusions are summarized as follows.\\

$\bullet$ We have reviewed the LLCD mechanism in the context of a MSP for accelerating  electrons to the UHE in the magnetosphere. A two-step process of electron acceleration is adopted: {\bf 1.} generation of electrostatic Langmuir waves from the conversion of star's rotational energy to the electrostatic energy and {\bf 2.} an efficient collapse of Langmuir waves on local beam of electrons leading to their acceleration.\\
\\
$\bullet$ If electrons reach at the level $\geq 0.01$ EeV in the young pulsar's magnetosphere, they interacting with targets/ambient matter (positrons) and radiation (photons) in the acceleration region just beyond the light cylinder of the MSP, produce PeV neutrinos and gamma rays.\\                                                     
\\ 
$\bullet$ In the model calculation, the level of diffuse neutrino energy flux per flavor contributed by all the newly born MSPs in the universe at redshifts in the range $0.002$--$5$ is found comparable to the total PeV neutrino flux per flavor observed by the IceCube. The observed diffuse flux of PeV neutrinos is useful to get an idea that how much fraction of the bolometric luminosity of a MSP that actually goes into UHE electrons. \\
\\
$\bullet$ Hadronic scenarios are widely accepted for the production of high-energy neutrinos. Most of the models proposed in earlier  studies took $\rm p\gamma$ and/or $\rm pp$ interactions for the production of UHE neutrinos in different sources. The present study involves the lepton-lepton interactions, occurring near the acceleration region of the young MSPs, generating the UHE astrophysical neutrinos. Here, we have shown that the leptonic scenario in some extreme astrophysical settings, such as newly born extragalactic MSPs, might have contributed to the PeV neutrino flux observed by the IceCube experiment.\\ 

$\bullet$ We have derived the associated gamma ray flux from the theory, and the amount of the flux is same as the neutrino flux. These extragalactic diffuse PeV gamma rays are unlikely to be available on earth due to stronger absorption in regions surrounding their production site, and over their long cosmological distances. A fraction of these PeV gamma rays might be converted into TeV gamma rays over the cosmological distance and would be detected at gamma ray observatories. 

In the work, we have estimated the diffusive flux of PeV scale neutrinos contributed by all the extragalactic MSPs. We have then concluded that a certain percentage ($\sim 0.31\%$) of the bolometric luminosity of these sources is necessary to produce the IceCube's detected neutrino flux of PeV scale if these objects were the only emitters of PeV neutrinos. 
We have not focused on the issue that how much percentage of the observed PeV neutrino flux was contributed by these MSPs in a situation where other type of objects (AGNs/Blazars..etc) might have produced neutrinos. Again, for different XLFs, exploited in the model calculation, the required percentages of the bolometric luminosity to power the IceCube PeV energy neutrinos are different. In  [60], it was shown that the blazar neutrinos mainly contributed by FRSQs, can only contribute to maximum $\sim 10\%$ of the IceCube detected flux via  $p\gamma$ interactions using the Lambda CDM cosmological framework. While in a different study [8], it was claimed that an acceptably small fraction $\sim 0.003\%$ of the bolometric luminosity of AGNs was adequate to power the observed diffuse fluxes of extragalactic PeV scale neutrinos. In their work they did consider neither the AGN cosmological evolution nor the Lambda CDM cosmological framework. Employing the Lambda CDM cosmological framework, we have carried out a calculation for obtaining the upper limit of the diffuse fluxes of neutrinos from AGNs via $p\gamma$ interactions in a different study. It came around $0.2\%$ of the AGN bolometric luminosity, and the work is still being under reviewed. An updated analysis of IceCube's 7.5 years data (2010-2017) has recently reported the detection of a cascade of high-energy particles (a particle shower), consistent with being created at the Glashow resonance with measured energy $6.05\pm 0.72$~PeV [64]. Their best fitted diffuse neutrino energy flux per flavor takes the form, $\Phi_{\rm{best-fit}}\approx 1.9\times 10^{-9}$ ~GeV~cm$^{-2}$~s$^{-1}$~sr$^{-1}$, which does have noticeable affect to our model predictions.     

The statistical uncertainties to all the best-fit XLF parameters are negligibly small (up to the order of $\approx \pm {3}~$\%) [50]. We ascertain a very little effect of these uncertainties on the estimated fluxes.  

%**************************************************************
\begin{acknowledgements}
	RKD acknowledges the financial support from North Bengal University under the Teachers' Research Project Scheme; Ref.No. 1513/R-2020. 
\end{acknowledgements}

\end{document}